\documentclass[preprint,showkeys,showpacs]{revtex4}
\usepackage{graphicx}
\usepackage{dcolumn}
\usepackage{amsmath}
\usepackage{color}
\usepackage{epstopdf}
\usepackage{graphicx}
\usepackage{subfigure}
\usepackage[utf8x]{inputenc}
\definecolor{coolblack}{rgb}{0.0, 0.18, 0.39}
\definecolor{darkred}{rgb}{0.5,0,0}
\definecolor{darkgreen}{rgb}{0,0.5,0}
\definecolor{darkblue}{rgb}{0,0,0.5}
\definecolor{lapislazuli}{rgb}{0.15, 0.38, 0.61}
\definecolor{venetianred}{rgb}{0.78, 0.03, 0.08}
\definecolor{bleudefrance}{rgb}{0.19, 0.55, 0.91}
\definecolor{dogwoodrose}{rgb}{0.84, 0.09, 0.41}
\usepackage[linktocpage,colorlinks]{hyperref}
\hypersetup{colorlinks=true, citecolor=darkgreen, linkcolor=blue,urlcolor = darkblue}
\makeatletter
\def\btt#1{\texttt{\@backslashchar#1}}
\DeclareRobustCommand\bblash{\btt{\@backslashchar}} \makeatother

\RequirePackage{fix-cm}

\begin{document}
\title{Stability Analysis of Geodesics and Quasinormal Modes of a Dual 
	Stringy Black Hole Via Lyapunov Exponents
}
\author{Shobhit Giri $^{a}$}\email{shobhit6794@gmail.com}
\author{ Hemwati Nandan $^{a,b}$}\email{hnandan@associates.iucaa.in}
\affiliation{$^{a}$Department of Physics,Gurukula Kangri (Deemed to be University), Haridwar 249 404, Uttarakhand, India}
\affiliation{$^{b}$Center for Space Research, North-West University, Mahikeng 2745, South Africa}


\begin{abstract}
 \noindent We investigate the stability of both timelike as well as null circular geodesics in the vicinity of a dual (3+1) dimensional stringy black hole (BH) spacetime by using an excellent tool so-called Lyapunov exponent. The proper time ($\tau$) Lyapunov exponent ($\lambda_{p}$) and coordinate time ($t$) Lyapunov exponent~($\lambda_{c}$) are explicitly derived to analyze the stability of equatorial circular geodesics for the stringy BH spacetime with \emph{electric charge} parameter ($\alpha$) and \emph{magnetic charge} parameter~($Q$). By computing these exponents for both the cases of BH spacetime, it is observed that the coordinate time Lyapunov exponent of magnetically charged stringy BH for both timelike and null geodesics are independent of magnetic charge parameter $(Q)$. The variation of the ratio of Lyapunov exponents with radius of timelike circular orbits ($r_{0}/M$) for both the cases of stringy BH are presented. The behavior of instability exponent for null circular geodesics with respect to charge parameters ($\alpha$ and $Q$) are also observed for both the cases of BH.  Further, by establishing a relation between quasinormal modes (QNMs) and parameters related to null circular geodesics (like angular frequency and Lyapunov exponent), we deduced the QNMs (or QNM frequencies) for a massless scalar field perturbation around \emph{both} the cases of stringy BH spacetime in the eikonal limit. The variation of scalar field potential with charge parameters and angular momentum of perturbation ($l$)  are visually presented and discussed accordingly. 
\end{abstract}
 
 

\maketitle

\section{Introduction}
\noindent General relativity (GR) proposed by Einstein in 1915 which defines the gravity as a manifestation of the curvature of spacetime in four dimensions has successfully passed all the tests of gravity to confirm the predictions made by it at large scales in the universe over the period of a century \cite{Hartle2003,Wald1984c,chandrasekhar1983mathematical,joshi1993global,poisson2004relativist}. However, in view of quantum structure of a spacetime where gravity is extravagantly feeble and the unification of gravity with other fundamental forces of nature, alternative theories of gravity like the string theory have emerged to understand the quantum aspects of gravity which require the presence of extra dimensions \cite{polchinski1998string,sen1985equations,sen1992rotating}. There exists several black hole (BH) spacetimes in GR and alternative theories of gravity as a solution to the Einstein field equations \cite{Hartle2003,chandrasekhar1983mathematical,polchinski1998string,bronnikov2016black,de2012black}. The geodesics of test particles (timelike and null) in the background of such BH spacetimes have been widely studied in diverse context due to their astrophysical importance \cite{Hartle2003,chandrasekhar1983mathematical,flathmann2016analytic,chatterjee2019analytic,hackmann2010geodesic,uniyal2015geodesic,uniyal2017null}.   In general, the effects of the curvature in a spacetime geometry are discussed through the geodesic motion which further investigate the characteristics of the BH spacetime. The equatorial circular geodesics play important role in viewpoint of stability analysis of orbits among different kind of geodesics \cite{Hartle2003,chandrasekhar1983mathematical,poisson2004relativist}. 

A circular geodesic may either be stable or unstable and its stability can be analyzed by using  the effective potential approach where the effective potential shows a maxima corresponding to an unstable circular geodesic (see \cite{thorne2000gravitation} for details).
Particularly, the stability analysis of geodesics via Lyapunov exponent is one of the basic tool to link between non-linear Einstein’s GR and non-linear 
dynamics~\cite{pradhan2013lyapunov,lyapunov1992general,skokos2010lyapunov,sano1985measurement,sota1996chaos}.
Basically, the Lyapunov exponent (or Lyapunov characteristic exponent) of a dynamical system is a measure of the average rate of separation of nearby trajectories in phase space. If two nearby geodesics diverge then Lyapunov exponent should have a positive value and if they converge then Lyapunov exponent has a negative value~\cite{cardoso2009geodesic,sharif2017particle,pradhan2016stability,mondal2020geodesic,pradhan2015stability,jamil2015dynamics}. 
So far, the instability of unstable circular orbits around any BH spacetime can be quantified by a positive Lyapunov exponent as a consequence of the non-linearity of GR~\cite{cornish2003lyapunov,cornish2001chaos}.

The Lyapunov exponents are applicable to analyze the stability of both the regular and chaotic motion. Specifically, in the vacuum BH background in GR, the geodesic motion is generally regular i.e. the geodesics around the BH spacetime can not be chaotic but chaos itself is likely to develop along the unstable circular orbits under perturbation like the spin of a BH or spin of a test particle~\cite{hilborn2000chaos,Suzuki1997}. However, around a magnetized BH it becomes to be fully chaotic, except the case of some islands of regularity \cite{universe6020026,li2019chaotic,kolovs2017possible,stuchlik2016acceleration}. In particular, the harmonic oscillatory motion of charged test particles around stable circular geodesics is described by the perturbation of the equations of motion around these circular geodesics. In fact, a charged test particle starts oscillating around a stable circular geodesics located at the equatorial plane if it is slightly displaced from the equilibrium position (as discussed in \cite{kolovs2017possible}). However,  the dynamics of the charged test particles in background of a BH immersed into an asymptotically uniform magnetic field (i.e. as of a magnetized BH spacetime \cite{horowitz1992dark,dasgupta2009kinematics}) has been investigated by Stuchlík et al. \cite{stuchlik2016acceleration}. The chaotic scattering in the effective potential corresponding to the gravitational field of a BH in presence of the uniform magnetic field is analyzed by the Hamiltonian formalism of the charged particle dynamics \cite{stuchlik2016acceleration} which is governed by the non-linear equations of motion leading to the chaotic motion and existence of off-equatorial circular geodesics.  Moreover, the motion of the neutral test particles around a Schwarzschild BH (SBH) immersed in an external uniform magnetic field can be chaotic whether they have no interactions of electromagnetic forces \cite{li2019chaotic}.

The main objective of this work is to investigate the stability of circular geodesics (timelike and null) and QNMs by calculating the Lyapunov exponents for a well-known dual stringy BH spacetime following the approach as described in \cite{pradhan2016stability}. Garfinkle, Horowitz and Strominger (GHS) \cite{garfinkle1992erratum} have obtained the asymptotically flat solutions from dilaton–Maxwell–Einstein field equations in the context of string theory representing electric and dual magnetic BHs so-called stringy BHs~\cite{horowitz1992dark,dasgupta2009kinematics,kar1997stringy,Uniyal:2014oaa,kuniyal2016null}. 
The spacetime geometry of these two BHs are quite similar to geometry of the SBH. 
Here, we also compared the all results obtained for GHS electric and magnetic (dual) solutions with those for SBH (by setting the stringy parameters to zero).\\
The analysis of geodesics stability around various BH spacetimes using Lyapunov exponent has already been extensively investigated in many articles. The existence and stability of circular geodesics in the background of Reissner-Nordstr\"{o}m spacetime were examined in detail \cite{pradhan2016stability} and the analysis of the Kolmogorov-Sinai (KS) entropy, innermost stable circular orbits (ISCO) for Kerr-Newman BH spacetime has also been performed via Lyapunov exponents \cite{pradhan2012isco}. The instability of both timelike and null circular geodesics in the equatorial plane for charged Myers Perry BH spacetimes has been widely investigated by P. Pradhan \cite{pradhan2013lyapunov}. Recently, geodesic stability and QNMs via Lyapunov exponent for a Hayward BH are analyzed \cite{mondal2020geodesic}.\\
Our approach here is to discuss the stability of geodesics via analyzing the both \emph{proper time and coordinate time} Lyapunov exponents. We further establish a relation between Lyapunov exponent (reciprocal of instability time scale of orbits) and QNMs of unstable null circular geodesics for electric and magnetic charged metrics. Null circular geodesics plays a crucial role to describe the characteristic modes of a BH, so-called QNMs which can be interpreted as null particles trapped at the unstable circular 
orbit~\cite{berti2009quasinormal,kokkotas1999quasi,nollert1999quasinormal,prasia2017quasinormal,konoplya2011quasinormal}. 
However, Cardoso et al. \cite{cardoso2009geodesic} has already investigated that at the eikonal approximation, the real part of the complex QNMs of spherically symmetric, asymptotically flat spacetime is defined by the angular frequency and the imaginary part is related to the instability timescale of the orbit (i.e. Lyapunov exponent) of the unstable null circular geodesics.\\
The relation between the QNMs in eikonal approximation and the properties in unstable circular null geodesics hold in GR, but it is not universal for any gravity model such as the
gravitational perturbations of BHs in the Einstein-Lovelock theory \cite{konoplya2017eikonal,stuchlik2019shadow}. More generally, the geodesic analogy is likely to break down in the generic case of coupling between tensor and scalar fields in modified theories of gravity. Harko et al. \cite{harko2014generalized} has extensively studied the curvature-matter couplings in f(R)-type modified gravity models inducing the covariant derivative of the energy-momentum tensor. In view of such coupling, the test particles follow a non-geodesic path which leads to the appearance of an extra force orthogonal to the four-velocity. Moreover, it could be violated even in the case of electromagnetic perturbations of BHs in non-linear electrodynamics as recently studied by Toshmatov et al. \cite{toshmatov2019relaxations}.
The relationship between Lyapunov exponent and radial effective potential is derived (see \textbf{Appendix-A} for details) for the stability analysis of a given BH spacetime.

\subsection{The Critical Exponent}
Let us define the critical exponent ($\gamma$), a quantitative characterization of instability of
circular geodesics by introducing a typical orbital timescale $T_{\Omega}=\frac{2\pi}{\Omega}$ and 
Lyapunov time scale or instability time scale $T_{\lambda}=\frac{1}{\lambda}$ 
as~\cite{pretorius2007black,cardoso2009geodesic,pradhan2016stability},
\begin{equation}
	\gamma= \frac{T_{\lambda}}{T_{\Omega}}= \frac{\Omega}{2\pi \lambda}\label{eq21}~.
\end{equation}
The relations between critical exponent and second order derivative of the radial 
effective potential i.e. $(\dot{r}^{2})^{''}$ corresponding to proper time and coordinate time are expressed as follows,
\begin{equation}
	\gamma_{p}= \frac{\Omega}{2\pi \lambda_{p}}=\frac{1}{2\pi}\sqrt{\frac{2\Omega^{2}}{(\dot{r}^{2})^{''}}},~~~
	\gamma_{c}= \frac{\Omega}{2\pi \lambda_{c}}=\frac{1}{2\pi}\sqrt{\frac{2\dot{\phi}^{2}}{(\dot{r}^{2})^{''}}},\label{eq22}
\end{equation}
where, $\Omega$ is angular frequency or orbital angular velocity.\\
For the expressions of proper time and coordinate time Lyapunov exponents i.e. $\lambda_{p}$ and $\lambda_{c}$ as mentioned in Eq.(\ref{eq22}) (see Eqs.(\ref{eq19}) and (\ref{eq20})  in Appendix-A).
Here, in our investigation, we will focus to determine critical exponent to confirm the observational relevance of instability of equatorial circular geodesics (for timelike and null) for the dual stringy BH spacetimes.
\section{ Lyapunov Exponents and Geodesic Stability of Stringy BH with Electric Charge }
In this section, we consider a static, spherically symmetric spacetime of electric charged BH 
emerging out of string theory in dilaton–Maxwell gravity. We quote the metric of the BH with electric charge as \cite{horowitz1992dark,dasgupta2009kinematics},
\begin{equation}
	dS_{Ele}^{2}= -\frac{\left(1-\frac{2M}{r}\right)}{\left(1+\frac{2M \sinh^{2}\alpha}{r}\right)^{2}} dt^{2}+\frac{d r^{2}}{\left(1-\frac{2M}{r}\right)} + r^{2} d\Omega_{2}^{2}~,\label{eq23}
\end{equation}
where, $d\Omega_{2}^{2}= \left(d\theta^{2}+\sin^{2}\theta d\phi^{2}\right)$, is the  metric for a 2-dimensional unit sphere and $\alpha$ is electric charge parameter. \\

In order to investigate the geodesics of test particles in the equatorial plane for above mentioned spacetime, we have to determine the Lagrangian for the motion by setting $\theta=\frac{\pi}{2}$.
The necessary Lagrangian for test particle motion in the equatorial plane can be written as \cite{Hartle2003},
\begin{equation}
	2\mathcal{L}_{Ele}= -\frac{\left(1-\frac{2M}{r}\right)}{\left(1+\frac{2M \sinh^{2}\alpha}{r}\right)^{2}} 
	\dot{t}^{2}+\frac{\dot{r}^{2}}{\left(1-\frac{2M}{r}\right)} + r^{2} \dot{\phi}^{2}. \label{eq24}
\end{equation}
Since the spacetime is static and spherically symmetric as independent of coordinates $t$ and $\phi$.  Therefore, the generalized momenta corresponding to these coordinates  will produced the two constants of motion so-called energy~($E$) and angular momentum~($L$) per unit rest mass of the particle as below,
\begin{equation}
	p_{t}=-\frac{\left(1-\frac{2M}{r}\right)}{\left(1+\frac{2M \sinh^{2}\alpha}{r}\right)^{2}} \dot{t}
	=-E ,\label{eq25}
\end{equation}

\begin{equation}
	p_{\phi}=r^{2}\dot{\phi}=L .\label{eq26}
\end{equation}
From above Eqs.(\ref{eq25}) and (\ref{eq26}), we deduce,
\begin{equation}
	\dot{t}=\frac{E \left(1+\frac{2M \sinh^{2}\alpha}{r}\right)^{2}}{\left(1-\frac{2M}{r}\right)} 
	,~~~~ \dot{\phi}= \frac{L}{r^{2}}.\label{eq27}
\end{equation}

The motion of the test particle around any BH spacetime is constrained as,
\begin{equation}
	g_{\mu \nu} \dot{x}^{\mu}\dot{x}^{\nu}=\delta ,\label{eq28}
\end{equation}
which leads to,

\begin{equation}
	-\frac{\left(1-\frac{2M}{r}\right)}{\left(1+\frac{2M \sinh^{2}\alpha}{r}\right)^{2}} 
	\dot{t}^{2}+\frac{\dot{r}^{2}}{\left(1-\frac{2M}{r}\right)} +r^{2} \dot{\phi}^{2}
	=\delta,\label{eq29}
\end{equation}\\
here and throughout this paper, for metric signature $(-,+,+,+)$, one can set $\delta$ = 0, -1 and +1 corresponding to null, timelike and spacelike geodesics respectively.\\

Now, by substituting $\dot{t}$ and $\dot{\phi}$ from Eq.(\ref{eq27}) into Eq.(\ref{eq29}), the radial equation for stringy BH with electric charge is expressed as,
\begin{equation}
	\dot{r}^{2}= E^{2} \left(1+\frac{2M \sinh^{2}\alpha}{r}\right)^{2}-\left(1-\frac{2M}{r}\right) 
	\left(\frac{L^{2}}{r^{2}}-\delta \right).\label{eq30}
\end{equation}

\subsection{For Timelike Circular Geodesics ($\delta =-1$)}
For timelike geodesics of test particle, the radial equation~(\ref{eq30}) with $\delta=-1$ leads to,
\begin{equation}
	\dot{r}^{2}= E^{2} \left(1+\frac{2M \sinh^{2}\alpha}{r}\right)^{2}-\left(1+\frac{L^{2}}{r^{2}}\right)
	\left(1-\frac{2M}{r}\right).\label{eq31}
\end{equation}
However, in order to restrict the motion of the test particle to circular geodesic motion in the gravitational field of specified BH, one may consider the constant radius ($r = r_{0}$) of orbits.

Thus, from the radial equation (\ref{eq31}), when the condition (\ref{eq18}) for the occurrence of circular orbits is taken into account, the energy per unit mass of the test particle is obtained as,

\begin{equation}
	E_{0}^{2}= \frac{r_{0} (2M-r_{0})^{2}}{r_{0}^{3}-3M r_{0}^{2}-8M^{2}r_{0}\sinh^{2}\alpha
		+2Mr_{0}^{2}\sinh^{2}\alpha-4M^{3}\sinh^{4}\alpha}.\label{eq32}
\end{equation}

However, the angular momentum per unit mass of particle is calculated as,

\begin{equation}
	L_{0}^{2}= \frac{r_{0}^2 (Mr_{0}+2M r_{0} \sinh^{2}\alpha-2M^{2}\sinh^{2}\alpha )}{r_{0}^{2}-3M r_{0}-2M^{2} 
		\sinh^{2}\alpha}.\label{eq33}
\end{equation}
The energy and angular momentum must be real and finite for the existence of circular motion, therefore the following three conditions,

\begin{equation}
	r_{0}^{3}-3M r_{0}^{2}-8M^{2}r_{0}\sinh^{2}\alpha+2Mr_{0}^{2}\sinh^{2}\alpha -4M^{3}\sinh^{4}\alpha>0,
\end{equation}

\begin{equation}
	Mr_{0}+2M r_{0} \sinh^{2}\alpha-2M^{2}\sinh^{2}\alpha>0 ,
\end{equation}

\begin{equation}
	r_{0}^{2}-3M r_{0}-2M^{2} \sinh^{2}\alpha>0,
\end{equation}
must be satisfied simultaneously.

For convenience, one can introduce some new quantities associated with above expressions which reads as,

\begin{equation}
	r_{0}^{3}-3M r_{0}^{2}-8M^{2}r_{0}\sinh^{2}\alpha+2Mr_{0}^{2}\sinh^{2}\alpha -4M^{3}\sinh^{4}\alpha =X,
\end{equation}

\begin{equation}
	r_{0}^{2}-3M r_{0}-2M^{2} \sinh^{2}\alpha=Y,
\end{equation}

\begin{equation} 
	Mr_{0}+2M r_{0} \sinh^{2}\alpha-2M^{2}\sinh^{2}\alpha=Z.
\end{equation}

We further evaluated an important quantity associated with the timelike circular geodesics at $r=r_{0}$ i.e. so-called orbital angular velocity or angular frequency ($\Omega_{0}$) as,

\begin{multline}
	\Omega_{0}^{Ele}=\frac{\dot{\phi}}{\dot{t}} = \left(\frac{(Mr_{0}+2M r_{0} \sinh^{2}\alpha-2M^{2}\sinh^{2}\alpha)}{r_{0}(r_{0}+2M \sinh^{2}\alpha)^{4}}\right)^{1/2} \\\left(\frac{(r_{0}^{3}-3M r_{0}^{2}-8M^{2}r_{0}\sinh^{2}\alpha+2Mr_{0}^{2}\sinh^{2}\alpha-4M^{3}\sinh^{4}\alpha)}{(r_{0}^{2}-3M r_{0}-2M^{2}\sinh^{2}\alpha)}\right)^{1/2}  ,\label{eq34}
\end{multline}

which can be rewritten in the following form,

\begin{equation}
	\Omega_{0}^{Ele}=\sqrt{\frac{Z X}{r_{0}(r_{0}+2M \sinh^{2}\alpha)^{4}Y}}.\label{eq35}
\end{equation}
The proper time Lyapunov exponent for electric charged stringy BH by using Eq.(\ref{eq19}) is then derived as,
\begin{multline}
	\lambda_{p}^{Ele}= \\ \left[\frac{1}{r_{0}^{3}} \left(\frac{ (12M-3r_{0})Z}{Y}+2M+\frac{(2M-r_{0})^{2}(12M^{2}\sinh^{4}\alpha+4Mr_{0}\sinh^{2}\alpha)}{X}\right)\right]^{1/2},\label{eq36}
\end{multline}

and from Eq.(\ref{eq20}), the coordinate time Lyapunov exponent for electric charged stringy BH is also obtained as follows,

\begin{multline}
	\lambda_{c}^{Ele} =  \left(\frac{X}{r_{0}^{2}(r_{0}+2M \sinh^{2}\alpha)^{4}}\right)^{1/2}\\\left(\frac{(12M-3r_{0})Z}{Y}+2M
	+\frac{(2M-r_{0})^{2}(12M^{2}\sinh^{4}\alpha+4Mr_{0}\sinh^{2}\alpha)}{X}\right)^{1/2} .\label{eq37}
\end{multline}

Let us introduce the bracketed term as mentioned in the above two expressions by a new quantity defined as below,
\begin{equation}
	\Delta = \left(\frac{(12M-3r_{0})Z}{Y}+2M+\frac{(2M-r_{0})^{2}(12M^{2}\sinh^{4}\alpha
		+4Mr_{0}\sinh^{2}\alpha)}{X}\right).
\end{equation}

Hence, the above two expressions of Lyapunov exponents reduce to the following simple forms,

\begin{equation}
	\lambda_{p}^{Ele}= \sqrt{\frac{\Delta}{r_{0}^{3}}} ,\label{eq38}
\end{equation}

\begin{equation} 
	\lambda_{c}^{Ele}= \sqrt{\frac{X \Delta}{r_{0}^{2}(r_{0}+2M \sinh^{2}\alpha)^{4}}} .\label{eq39}
\end{equation}

\noindent We now illustrate the stability of the timelike circular geodesics for stringy BH with electric charge via analyzing the nature of the both Lyapunov exponents as follows:

\noindent The timelike circular geodesics are stable when both exponents $\lambda_{p}^{Ele}$ and 
$\lambda_{c}^{Ele}$ are imaginary i.e. $\Delta< 0$ while the circular geodesics are unstable when both Lyapunov exponents are real, for which $\Delta> 0$. The timelike circular geodesic are marginally stable when $\Delta=0$ so that both $\lambda_{p}^{Ele}$ and $\lambda_{c}^{Ele}$ vanish simultaneously.  \\

The critical exponents for electric charged stringy BH are calculated as,

\begin{equation}
	\gamma_{p}^{Ele}=\frac{\Omega_{0}^{Ele}}{2\pi\lambda_{p}^{Ele}}
	=\frac{1}{2\pi}\sqrt{\frac{Z X r_{0}^{2}}{(r_{0}+2M \sinh^{2}\alpha)^{4} Y \Delta}},\label{eq40}
\end{equation}

\begin{equation}
	\gamma_{c}^{Ele}=\frac{\Omega_{0}^{Ele}}{2\pi\lambda_{c}^{Elec}}=\frac{1}{2\pi}\sqrt{\frac{Z r_{0}}{Y \Delta}} 
	.\label{eq41}
\end{equation}

Hence, for any unstable circular orbit, the quantity $\Delta>0$ such that the Lyapunov time scale is shorter than the gravitational time scale (i.e. $T_{\lambda}<T_{\Omega}$) which confirms observational relevance of the instability of circular geodesics.

\noindent Besides this, one can express the ratio of Lyapunov exponent $\lambda_{p}^{Ele}$ to $\lambda_{c}^{Ele}$ as given below,

\begin{equation}
	\left(\frac{\lambda_{p}}{\lambda_{c}}\right)^{Ele} = \frac{(r_{0}+2M \sinh^{2}\alpha)^{2}}{\sqrt{r_{0}X}}.\label{eq42}
\end{equation}

\begin{figure}[]
	\centering
	\includegraphics[width=9cm,height=09cm]{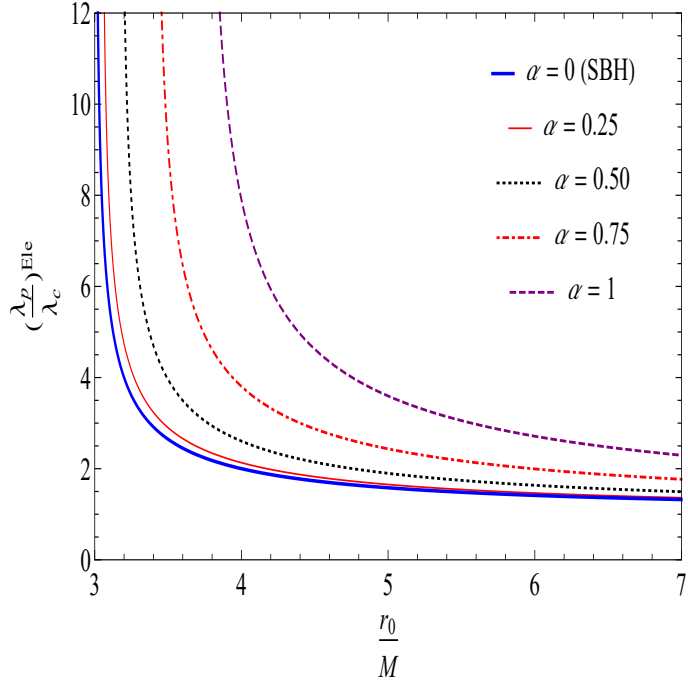}
	\caption {Comparison of the variation of ratio ``$(\frac{\lambda_{p}}{\lambda_{c}})^{Ele}$" with ``$r_{0}/M$" of electric charged stringy BH for different values of electric charge parameter $\alpha$ and SBH case (i.e. $\alpha=0$).}\label{LEratioEle}
\end{figure}
The comparison of the variation of the ratio of both the Lyapunov exponents as a function of the radius of circular orbit to mass ratio $(r_{0}/M)$ for electric charged stringy BH with various values of electric charge parameter ($\alpha$) and for SBH case ( i.e. $\alpha=0$) can be visualized in \figurename{\ref{LEratioEle}}. The range of radius of circular orbit is fixed within the null circular orbit radius (which is larger than $r_{c}=3M$ for SBH  ) and radius of innermost stable circular orbit (ISCO) (which is larger than $r_{ISCO}=6M$ for SBH). It is straight forward to conclude that for different values of electric charge parameter $\alpha$, the ratio $\left(\frac{\lambda_{p}}{\lambda_{c}}\right)^{Ele}$ decreases exponentially with radius $r_{0}/M$.\\

\subsection{For Null Circular Geodesics ($\delta=0$)}
For null geodesics, one can only calculate the coordinate time Lyapunov exponent($\lambda_{c}$) for both BH spacetimes since lightlike particles have no proper time. So, by replacing $\delta=0$ from Eq.(\ref{eq30}), the radial equation for null geodesic motion is described as,

\begin{equation}
	\dot{r}^{2}= E^{2} \left(1+\frac{2M \sinh^{2}\alpha}{r}\right)^{2}-\left(\frac{L^{2}}{r^{2}}\right)
	\left(1-\frac{2M}{r}\right).\label{eq43}
\end{equation}

Now by considering the condition (\ref{eq18}) for null circular geodesic at radius $r=r_{c}$, the ratio of energy to angular momentum is obtained as below,

\begin{equation}
	\frac{E_{c}}{L_{c}}(Null) = \pm \sqrt{\frac{r_{c}-2M}{r_{c}(r_{c}+2M \sinh^{2}\alpha)^{2}}} ,\label{eq44}
\end{equation}
\\
and a quadratic equation of radius $r_{c}$ is deduced as,

\begin{equation}
	r_{c}^{2}-3M r_{c}-2M^{2} \sinh^{2}\alpha=0.\label{eq45}
\end{equation}

By solving Eq.(\ref{eq45}), the radius of null circular orbit or photon sphere is determined as below,

\begin{equation}
	r_{c_{\pm}}= \frac{3M}{2}\left[1+\sqrt{1+\frac{8}{9}\sinh^2\alpha}\right] .\label{eq46}
\end{equation}

The angular frequency for null circular geodesics at $r=r_{c}$ is evaluated as,

\begin{equation}
	\Omega_{c}^{Ele}= \frac{\dot{\phi}}{\dot{t}}= \sqrt{\frac{r_{c}-2M}{r_{c}(r_{c}
			+2M \sinh^{2}\alpha)^{2}}}.\label{eq47}
\end{equation}
The impact parameter associated with the null circular geodesics is found accordingly as,\\
\begin{equation}
	D_{c}= \frac{L_{c}}{E_{c}}= \frac{1}{\Omega_{c}^{Ele}}= \sqrt{\frac{r_{c}(r_{c}
			+2M \sinh^{2}\alpha)^{2}}{r_{c}-2M}}.\label{eq48}
\end{equation}

The angular frequency of null circular geodesics is therefore defined as inverse of the impact parameter associated with it. Finally, from Eq.(\ref{eq20}), the Lyapunov exponent for null circular geodesics is derived as,

\begin{widetext}
	\begin{equation}
		\lambda_{Null}^{Ele}=\sqrt{\frac{\left(r_{c}-2M\right)^{2}}{U^{4}}\left[\frac{4M^{2}\sinh^{4}\alpha}{r_{c}^{2}}
			+\frac{4M \sinh^{2}\alpha U}{r_{c}^{2}}-\frac{3U^{2}}{r_{c}^{2}}
			+\frac{6M U^{2}}{r_{c}^{2}\left(r_{c}-2M\right)}\right]},\label{eq49}
	\end{equation}
	
\end{widetext}

\noindent where, the quantity $U=\left(r_{c}+2M\sinh^{2}\alpha\right)$.

Since the bracketed term in the above expression is positive always such that $\lambda_{Null}^{Ele}$ is real. Hence it may be concluded that null circular geodesics are unstable at the radius $r_{c_{\pm}}$. 
The Lyapunov exponent and angular frequency corresponding to the null circular geodesics are crucial to describe the instability of unstable circular orbits. One can therefore represent the instability exponent ($\lambda_{Null}/\Omega_{c}$) in the following form,

\begin{equation}
	\left(\frac{\lambda_{Null}}{\Omega_{c}}\right)^{Ele}= \sqrt{\frac{r_{c}(r_{c}-2M)\left[\frac{4M^{2}\sinh^{4}
				\alpha}{r_{c}^{2}}+\frac{4M \sinh^{2}\alpha U}{r_{c}^{2}}-\frac{3U^{2}}{r_{c}^{2}}
			+\frac{6M U^{2}}{r_{c}^{2}\left(r_{c}-2M\right)}\right]}{U^{3}}}.
\end{equation}\label{InstExp1}

\begin{figure}[]
	\centering
	\includegraphics[width=8cm,height=7cm]{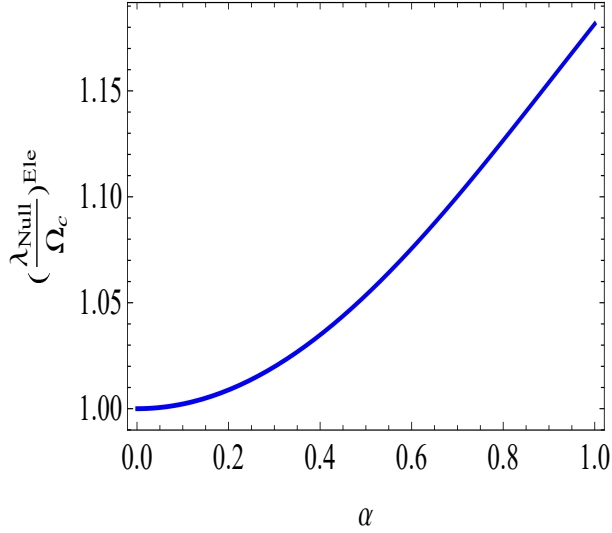}
	\caption{The behavior of ``$\left(\frac{\lambda_{Null}}{\Omega_{c}}\right)^{Ele}$"  for electric charged stringy BH with electric charge parameter ``$\alpha$". Here $r_{c}=3M$ (with $M=1$).}\label{LbyOele}
\end{figure}

The behavior of instability exponent ($\lambda_{Null}/\Omega_{c}$) with respect to electric charge parameter ($\alpha$) for radius of circular orbit $r_{c}=3M$  is presented in \figurename{\ref{LbyOele}}. One can observe that the instability exponent is increasing exponentially with increase in charge parameter $\alpha$.
\\
The critical exponent i.e. the quantitative characterization of null circular geodesics for electric charged stringy BH is then expressed in the following form,

\begin{multline}
	\gamma_{Null}^{Ele}= \frac{\Omega_{c}^{Ele}}{2\pi \lambda_{Null}^{Ele}}
	\\ =\frac{1}{2\pi} \sqrt{\frac{U^{3}}{r_{c}(r_{c}-2M)\left[\frac{4M^{2}\sinh^{4}
				\alpha}{r_{c}^{2}}+\frac{4M \sinh^{2}\alpha U}{r_{c}^{2}}-\frac{3U^{2}}{r_{c}^{2}}
			+\frac{6M U^{2}}{r_{c}^{2}\left(r_{c}-2M\right)}\right]}}.\label{eq50}
\end{multline}

The Lyapunov exponent $\lambda_{Null}^{Ele}$ is real for the present case such that $T_{\lambda}<T_{\Omega}$ which reveals the instability of the circular null geodesics observationally for electric charged stringy BH. For SBH ($\alpha=0$) with $r_{c}=3M$, the Lyapunov exponent for null circular geodesics also yields the real value given as,

\begin{equation}
	\lambda_{Null}^{SBH}=\frac{1}{3\sqrt{3}M},\label{eq51}
\end{equation}

which in tern proves the instability of the Schwarzschild photon sphere. 
\section{ Lyapunov Exponents and Geodesic Stability of Stringy BH with Magnetic Charge}
In this section, we will perform the same analysis (as in previous section) of stability for the circular geodesics (timelike and null) of the test particle around stringy BH with the magnetic charge. The line element for magnetic charged stringy BH is given by~\cite{horowitz1992dark,dasgupta2009kinematics},

\begin{equation}
	dS_{Mag}^{2}= -\frac{\left(1-\frac{2M}{r}\right)}{\left(1-\frac{Q^{2}}{2M r}\right)} dt^{2}
	+\frac{d r^{2}}{\left(1-\frac{2M}{r}\right)\left(1-\frac{Q^{2}}{2M r}\right)} + r^{2} d\Omega_{2}^{2},\label{eq52}
\end{equation}
where, $Q$ is the magnetic charge parameter.

The necessary Lagrangian for the motion in equatorial plane $(\theta=\frac{\pi}{2})$ around the above mentioned BH spacetime reads as,

\begin{equation}
	2\mathcal{L}_{Mag}= -\frac{\left(1-\frac{2M}{r}\right)}{\left(1-\frac{Q^{2}}{2M r}\right)} \dot{t}^{2}
	+\frac{\dot{r}^{2}}{\left(1-\frac{2M}{r}\right)\left(1-\frac{Q^{2}}{2M r}\right)} \\
	+ r^{2} \dot{\phi}^{2}.\label{eq53}
\end{equation}
Since the metric is also independent of coordinates $t$ and $\phi$, therefore, the two conserved quantities associated with generalized momenta are obtained as,

\begin{equation}
	p_{t} = -\frac{\left(1-\frac{2M}{r}\right)}{\left(1-\frac{Q^{2}}{2M r}\right)} ~\dot{t} = -E\label{eq54} ,
\end{equation}

\begin{equation}
	p_{\phi} = r^{2}\dot{\phi} = L.\label{eq55}
\end{equation}
From Eqs.(\ref{eq54}) and (\ref{eq55}),

\begin{equation}
	\dot{t}=\frac{E \left(1-\frac{Q^{2}}{2M r}\right)}{\left(1-\frac{2M}{r}\right)} ,~~~\label{eq56}
	\dot{\phi}= \frac{L}{r^{2}}.
\end{equation}
However, the constraint Eq.(\ref{eq28}) for the null geodesics yields,

\begin{equation}
	-\frac{\left(1-\frac{2M}{r}\right)}{\left(1-\frac{Q^{2}}{2M r}\right)} \dot{t}^{2}+\frac{\dot{r}^{2}}
	{\left(1-\frac{2M}{r}\right)\left(1-\frac{Q^{2}}{2M r}\right)} +r^{2} \dot{\phi}^{2}=\delta .\label{eq57}
\end{equation}
By inserting the Eq.(\ref{eq56}) into Eq.(\ref{eq57}), the radial equation of motion for stringy BH with magnetic charge parameter is deduced in the following form,

\begin{equation}
	\dot{r}^{2}= E^{2} \left(1-\frac{Q^{2}}{2M r} \right)^{2} - \left(\frac{L^{2}}{r^{2}} - \delta \right) \left(1-\frac{2M}{r}\right) \left(1-\frac{Q^{2}}{2M r}\right).\label{eq58}
\end{equation}

\subsection{For Timelike Case ($\delta=-1$)}
For timelike geodesics, substituting $\delta=-1$ into Eq.(\ref{eq58}), the radial equation reads as,

\begin{equation} 
	\dot{r}^{2}= E^{2} \left(1-\frac{Q^{2}}{2M r}\right)^{2}-\left(1+\frac{L^{2}}{r^{2}}\right)
	\left(1-\frac{2M}{r}\right) \left(1-\frac{Q^{2}}{2M r}\right).\label{eq59}
\end{equation}
Now, by using above radial equation and the conditions for circular orbits at constant radius $r=r_{0}$ given in Eq.(\ref{eq18}), the energy and angular momentum per unit mass of test particle read as,

\begin{equation}
	E_{0}^{2}= \frac{-2M \left(2M-r_{0}\right)^{2}}{\left(3M-r_{0}\right)\left(2M r_{0}-Q^{2}\right)} ~ \& ~ L_{0}^{2}
	= \frac{-M r_{0}^{2}}{3M-r_{0}}.\label{eq60}
\end{equation}
\\
The circular geodesics are possible when the conditions $ \left(3M-r_{0}\right)<0$ and $\left(2M r_{0}-Q^{2}\right)>0$ are satisfied simultaneously such that $E_{0}$ and $L_{0}$  are to be real and finite.\\
Therefore, the orbital angular velocity or angular frequency ($\Omega_{0}$) for timelike circular geodesics is deduced as,

\begin{equation}
	\Omega_{0}^{Mag}=\frac{\sqrt{2}M}{r_{0}\sqrt{2Mr_{0}-Q^{2}}}.\label{eq61}
\end{equation}
The proper time Lyapunov exponent for timelike circular geodesics of magnetic charged stringy BH is then evaluated as,

\begin{equation}
	\lambda_{p}^{Mag}= \sqrt{\frac{-\left(2Mr_{0}-Q^{2}\right)}{2 r_{0}^{4}}
		\left(\frac{r_{0}-6M}{r_{0}-3M} \right)}.\label{eq62}
\end{equation}

\noindent However, the coordinate time Lyapunov exponent is calculated as,

\begin{equation}
	\lambda_{c}^{Mag}= \sqrt{\frac{-M\left(r_{0}-6M\right)}{r_{0}^{4}}}.\label{eq63}
\end{equation}

It is interesting to note that the coordinate time Lyapunov exponent is independent of magnetic charge parameter $Q$ of the BH and is identical with the expression as that for SBH. 

\noindent Thus, it can be concluded that the timelike circular geodesics for magnetically charged 
stringy BH are stable when $\left(r_{0}-6M\right)>0$  and $\left(2Mr_{0}-Q^{2}\right)>0$ simultaneously such that $\lambda_{p}^{Mag}$ and $\lambda_{c}^{Mag}$ are imaginary. The geodesics are unstable when  $\left(r_{0}-6M\right)<0$ and $\left(2Mr_{0}-Q^{2}\right)<0$ simultaneously i.e. both Lyapunov exponents should have a real value.

\noindent The critical exponents for timelike circular geodesics of magnetic charged stringy BH  is determined as below,

\begin{equation}
	\gamma_{p}^{Mag}=\frac{\Omega_{0}^{Mag}}{2\pi\lambda_{p}^{Mag}}
	=\frac{1}{2\pi}\frac{2M r_{0}}{(2Mr_{0}-Q^{2})}\sqrt{-\left(\frac{r_{0}-3M}{r_{0}-6M}\right)},\label{eq64}
\end{equation}

\begin{equation}
	\gamma_{c}^{Mag}=\frac{\Omega_{0}^{Mag}}{2\pi\lambda_{c}^{Mag}}
	=\frac{M r_{0}}{2\pi}\sqrt{\frac{-2}{M(2Mr_{0}-Q^{2})(r_{0}-6M)}}.\label{eq65}
\end{equation}

\noindent Thus, for any unstable circular orbit as mentioned above, the Lyapunov time scale is shorter than the gravitational time scale (i.e. $T_{\lambda}<T_{\Omega}$). Which in tern indicates the observational relevance of unstable timelike circular orbits.  

\begin{figure}[]
	\centering
	\includegraphics[width=8cm,height=7cm]{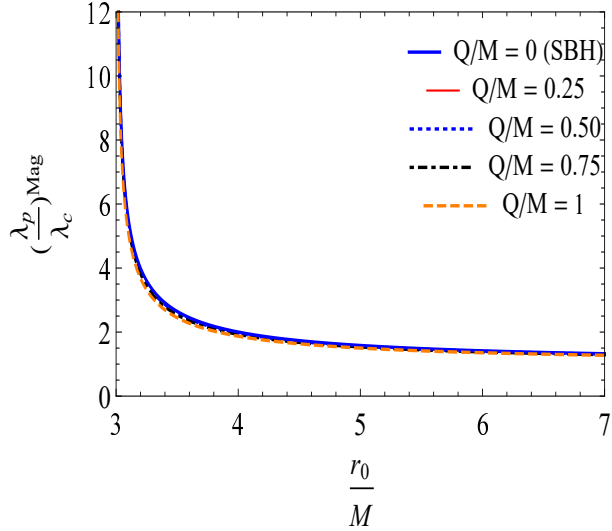}
	\caption {The variation of ratio ``$\left(\frac{\lambda_{p}}{\lambda_{c}}\right)^{Mag}$" with ``$r_{0}/M$" of magnetic charged stringy BH for various values of  parameter $Q/M$ and SBH ( i.e. $Q/M=0$).}\label{LEratioMag}
\end{figure}

However, we have also computed the ratio of proper time to coordinate time Lyapunov exponent for timelike circular geodesics as given below,

\begin{equation}
	\left(\frac{\lambda_{p}}{\lambda_{c}}\right)^{Mag}=\sqrt{\frac{\left(2M r_{0}-Q^{2}\right)}
		{2M \left(r_{0}-3M\right)}}.\label{eq66}
\end{equation}
From \figurename{\ref{LEratioMag}} , one can observe that the ratio of Lyapunov exponents as function of radius to mass ratio ($r_{0}/M$) with varying magnetic charge to mass ratio ($Q/M$) decreases exponentially in a way similar to the case of SBH ( i.e. $Q/M=0$).

By comparing Eq.(\ref{eq66}) with Eq.(\ref{eq42}), it can also be noticed that the contribution of electric charge parameter ($\alpha$) is more significant than the contribution of magnetic charge parameter ($Q$) which is clearly visualized in \figurename{\ref{LEratioEle}} and \figurename{\ref{LEratioMag}} respectively.

\subsection{For Null Case ($\delta=0$)}
In order to investigate null circular geodesics, substituting $\delta=0$ into Eq.(\ref{eq58}), the radial equation is obtained as follows,
\begin{equation} 
	\dot{r}^{2}= E^{2} \left(1-\frac{Q^{2}}{2M r}\right)^{2}-\left(\frac{L^{2}}{r^{2}}\right)
	\left(1-\frac{2M}{r}\right) \left(1-\frac{Q^{2}}{2M r}\right).\label{eq67}
\end{equation}
With the condition for circular null geodesics at radius $r=r_{c}$, one can find the ratio of energy and angular momentum of the particle as,

\begin{equation}
	\frac{E_{c}}{L_{c}}(Null) = \pm \sqrt{\frac{2M \left(r_{c}-2M\right)}{r_{c}^{2}
			\left(2M r_{c}-Q^{2}\right)}},\label{eq68}
\end{equation}

and the radius of null circular orbit or photon sphere is then given as,

\begin{equation}
	r_{c}=3M \hspace{0.2cm} or \hspace{0.2cm} \frac{Q^{2}}{2M}.\label{eq69}
\end{equation}
Here the choice $r_{c}=\frac{Q^{2}}{2M}$, is invalid because the above ratio is undefined in this limit. Therefore, $r_{c}=3M $ is only considerable choice. \\

\noindent The angular frequency measured by an asymptotic observer for null geodesics at $r=r_{c}$ is expressed as,

\begin{equation}
	\Omega_{c}^{Mag}= \frac{\dot{\phi}}{\dot{t}} = 
	\sqrt{\frac{2M \left(r_{c}-2M\right)}{r_{c}^{2}\left(2M r_{c}-Q^{2}\right)}},\label{eq70}
\end{equation}
and the impact parameter is given by,

\begin{equation}
	D_{c}= \frac{L_{c}}{E_{c}}= \frac{1}{\Omega_{c}^{Mag}}= 
	\sqrt{\frac{r_{c}^{2}\left(2M r_{c}-Q^{2}\right)}{2M \left(r_{c}-2M\right)}}
	.\label{eq71}
\end{equation}
So, the Lyapunov exponent (coordinate time) for null circular geodesics of stringy BH with magnetic charge is obtained as,

\begin{equation}
	\lambda_{Null}^{Mag}=\sqrt{\frac{\left(\dot{r}^{2}\right)^{''}}{2\dot{t}^{2}}}
	=\sqrt{\frac{3\left(r_{c}-2M\right)\left(4M-r_{c}\right)}{r_{c}^{4}}}.\label{eq72}
\end{equation} 

\begin{figure}[]
	\centering
	\includegraphics[width=8cm,height=7cm]{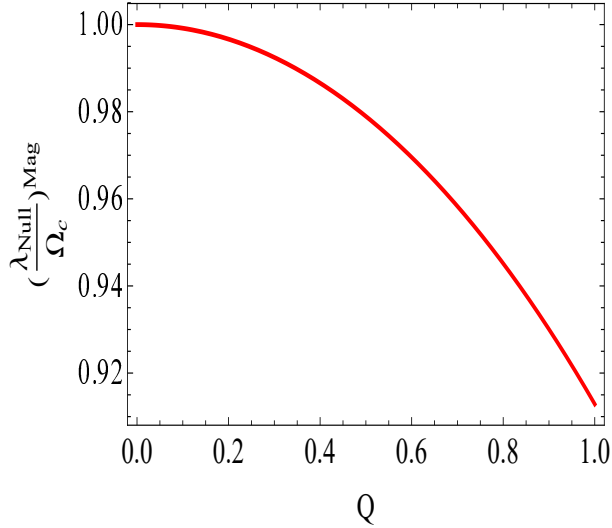}
	\caption{The behavior of ``$	(\frac{\lambda_{Null}}{\Omega_{c}})^{Mag}$"  for magnetic charged stringy BH as a function of magnetic charge parameter $Q$. Here $r_{c}=3M$ ( with $M=1$).}\label{LbyOmag}
\end{figure}

As a result, one can say that the Lyapunov exponent for null circular geodesics around magnetic charged stringy BH is independent of magnetic charge parameter ($Q$) and the null circular geodesics are unstable at $r_{c}=3M$ for which the value of $\lambda_{Null}^{Mag}=\frac{1}{3\sqrt{3}M}$ is real and finite.
The instability exponent ($\lambda_{Null}/\Omega_{c}$) of null circular orbits for stringy BH with magnetic charge is given by

\begin{equation}
	\left(\frac{\lambda_{Null}}{\Omega_{c}}\right)^{Mag}
	=\sqrt{\frac{3\left(2M r_{c}-Q^{2}\right)\left(4M-r_{c}\right)}
		{2M r_{c}^{2}}}.\label{InstExp2}
\end{equation}
The behavior of instability exponent ($\lambda_{Null}/\Omega_{c}$) with magnetic charge parameter ($Q$) for radius of circular orbit $r_{c}=3M$ for fixed value of mass $(M=1)$ is also presented in \figurename{\ref{LbyOmag}}. One can observe that the instability exponent is maximum for lowest values of magnetic charge parameter $Q$ and decreases sharply with increase in charge parameter $Q$.

Finally, one can determine the critical exponent for null circular geodesics of magnetic charged stringy BH as follow,
\begin{equation}
	\gamma_{Null}^{Mag}= \frac{\Omega_{c}^{Mag}}{2\pi \lambda_{Null}^{Mag}}
	=\frac{1}{2\pi}\sqrt{\frac{2M r_{c}^{2}}
		{3\left(2M r_{c}-Q^{2}\right)\left(4M-r_{c}\right)}}.\label{eq73}
\end{equation}
Since, the Lyapunov exponent $\lambda_{Null}^{Mag}$ is real and finite for magnetic charged  stringy BH i.e. the Lyapunov time scale is shorter than the gravitational time scale which confirms the instability of null circular geodesics in the present case.

\section{Quasinormal Modes (QNMs) for a Massless Scalar Field Perturbation around Stringy BHs Using Null Circular Geodesics}
In order to investigate the characteristic modes known as the QNMs at the eikonal limit of the 
stringy BH spacetimes, the parameters of the null circular geodesics play an important role in their determination \cite{cardoso2009geodesic}. S. Fernando has extensively studied QNM frequencies under massless scalar field perturbations near many BH spacetimes by using WKB approach 
and eikonal limit \cite{fernando2015regular,fernando2015quasi,fernando2017bardeen}.  
In this present study, the scalar perturbation by a massless field and the null 
geodesics around both charged BHs are to be employed to determine the QNMs at large angular momentum limit or eikonal limit.

The form of wave equation for such a field derived from the radial component of the Klein-Gordon equation will provide the QNM frequencies at the eikonal limit which can be simplified in following form \cite{cardoso2009geodesic},
\begin{equation}
	\frac{d^{2}\Psi(r_{*})}{dr^{2}_{*}}+P (r_{*})\Psi(r_{*})=0,\label{eq74}
\end{equation}
where,  $r_{*}$ is a “tortoise” coordinate (ranges from $-\infty$ at horizon to $+\infty$ at spatial infinity) and $P(r_{*})$ is defined as,
\begin{equation}
	P(r_{*})= \omega^{2}-V_{s}(r_{*}).\label{eq75}
\end{equation}
Here, $V_{s}(r_{*})$ is the scalar field potential.\\
The eikonal approximation is closely related to the WKB approximation which reduces the equations to differential equations with a single variable. For BH spacetimes having the wave equation of the form represented in Eq.(\ref{eq74}), the WKB methods result an accurate approximation of QNM frequencies in the eikonal limit.
Schutz and Will \cite{schutz1985black}  has developed a semianalytic technique for determining the complex normal mode frequencies or QNMs of BHs based on the WKB approximation. This technique in fact provides a simple analytic formula having the real and imaginary parts of the frequency in terms of the parameters of the BH, of the perturbation field and the quantity (n+1/2), where n = 0,1,2,... corresponding to the fundamental mode, first overtone and so on.  The general form of asymptotic WKB expansion at $r_{*} \rightarrow\pm\infty$ with the Taylor expansion as  discussed by Konoplya et al. \cite{konoplya2011quasinormal} is 
\begin{equation}
	\Psi(r_{*})= A (r_{*})~ exp\left(\sum_{n=0}^{\infty}\frac{S_{n}(r_{*})\epsilon^{n}}{\epsilon}\right),
\end{equation}
with the WKB parameter $\epsilon << 1$ to track orders of the WKB expansion.
For the effective potentials having a potential barrier and constant at the
event horizon and spatial infinity (i.e. $r_{*} \rightarrow\pm\infty$), three types of regions are discussed by  Schutz and Will \cite{schutz1985black} along with two turning points for which $P(r_{*})=0$. The Taylor series approximation leads to the solution of Eq.(\ref{eq74}) as below in region II when the turning points are closely spaced (i.e. $P(r_{*})_{max} >> P(\pm \infty) $), 
\begin{equation}
	P(r_{*})= P_{0}+ \frac{1}{2} \frac{d^{2}P_{0}}{dr_{*}^{2}}(r_{*}-r_{0})^{2} + O \left((r_{*}-r_{0})^{3}\right), \label{tyexp}
\end{equation}
where, $P_{0}$ represents the maximum of $P(r_{*})$ at a point $r_{*}=r_{0}$ for which $dP_{0}/d r_{*}=0$ and  $d^{2}P_{0}/dr_{*}^{2}$ is a second order derivative w.r.t. $r_{*}$ at that point.	However, corresponding to region II,
\begin{equation}
	|r_{*}-r_{0}| < \sqrt{\frac{-2P_{0}}{d^{2}P_{0}/dr_{*}^{2}}} \approx \epsilon^{1/2}.
\end{equation}
Therefore, the second order derivative term is negligible for $\epsilon<<1$ in above expansion given by Eq.(\ref{tyexp}). The QNMs condition from the wave equation (\ref{eq74}) is obtained as \cite{konoplya2011quasinormal,schutz1985black},

\begin{equation}
	P_{0}= i\left(n+\frac{1}{2}\right)\sqrt{2\frac{d^{2}P_{0}}{dr_{*}^{2}}},\label{eq79}
\end{equation}
where, $n$ is the overtone number. At a point $r_{0}=r_{c}$, the extremum $P_{0}$ and the location of the null circular geodesics must coincide.

The massless scalar field potential $V_{s}(r)$ in large $l$- limit i.e. $l \rightarrow \infty$ derived by Cardoso et al. \cite{cardoso2009geodesic} can be represented in the following form
\begin{equation}
	V_{s}(r)\approx l \frac{E_{c}^{2}}{L_{c}^{2}}=l \Omega_{c}^{2},\label{eq76}
\end{equation}
where, $l$ is the angular momentum of the perturbation.

Thus, the scalar field potential for case of electric charged stringy BH is expressed as
\begin{equation}
	V_{s}^{Ele}(r) = l (\Omega_{c}^{Ele})^{2}=l \frac{r_{c}-2M}{r_{c}(r_{c}+2M \sinh^{2}\alpha)^{2}}.\label{eq77}
\end{equation}

\begin{figure}[]
	\centering
	\subfigure[]{\includegraphics[width=8cm,height=7cm]{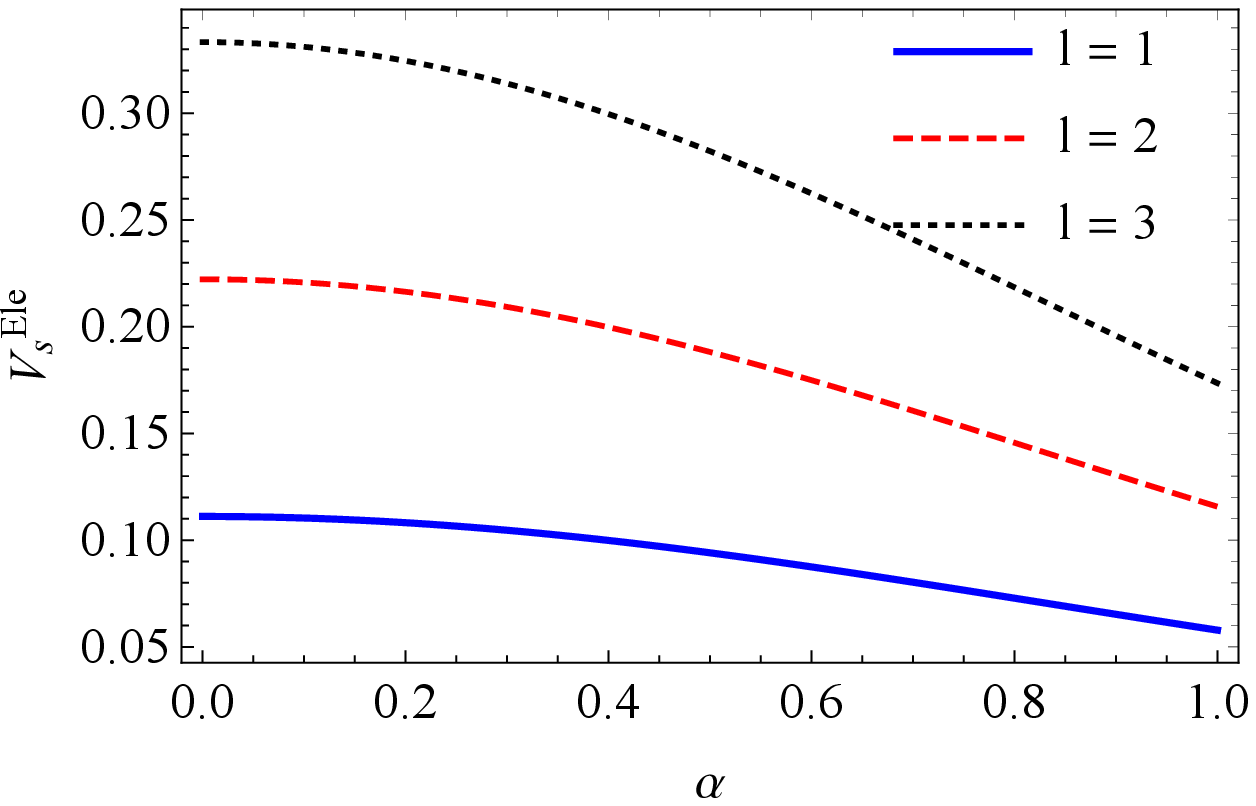}\label{Vselectrica}}
	\subfigure[]{\includegraphics[width=8cm,height=7cm]{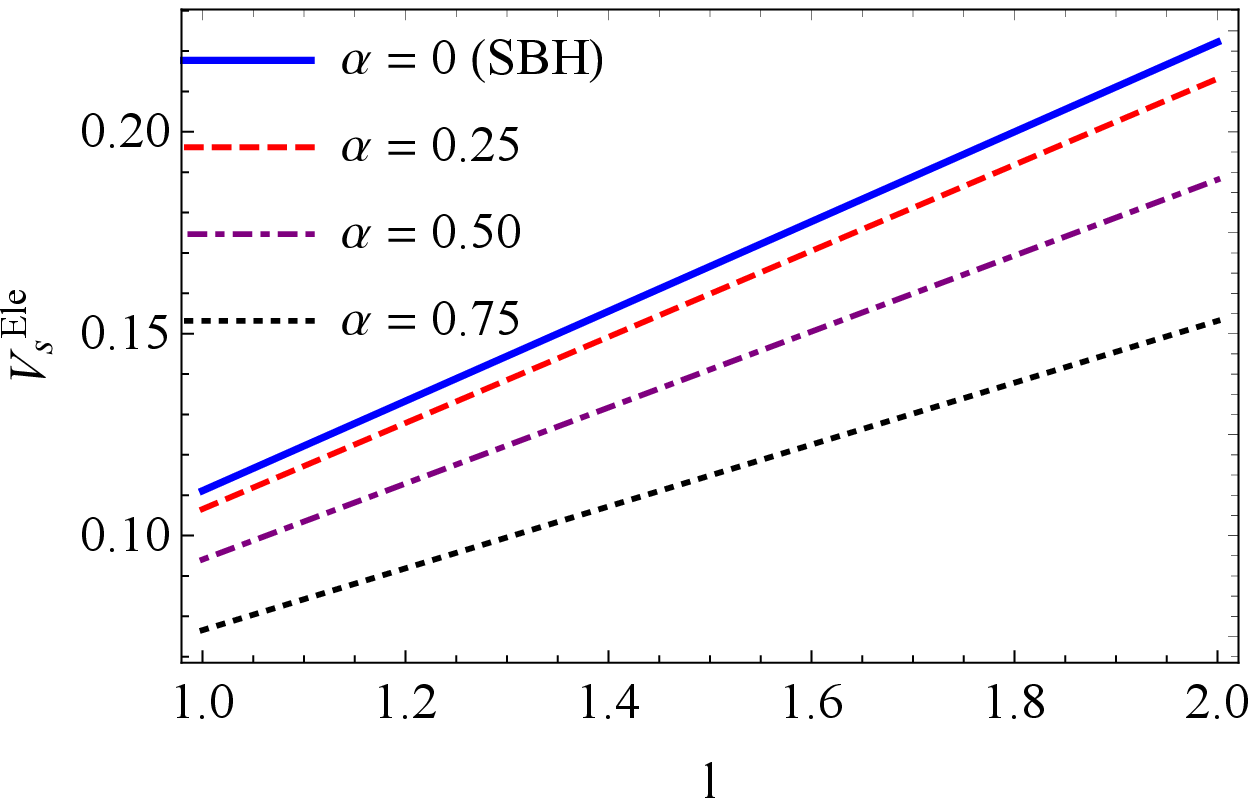}\label{Vselectricb}}
	\caption{The variation of scalar field potential $``V_{s}^{Ele}"$ with electric charge parameter $``\alpha"$ by varying the value of angular momentum of perturbation $(l)$ for fixed values $M=1$ and $r_{c}=3M$ (see upper panel). The variation of $``V_{s}^{Ele}"$ with $``l"$ by varying charge parameter $\alpha$ for fixed $M=1$ and $r_{c}=3M$ (see lower panel).}
\end{figure}

In this case, the scalar field potential depends on the mass of BH ($M$), electric charge parameter ($\alpha$) and angular momentum of the perturbation ($l$). In \figurename{\ref{Vselectrica}}, we have presented the variation of scalar potential as a function of electric charge parameter $\alpha$ by varying $l$ for fixed values of remaining parameters. It is observed that the potential decreases as values of $\alpha$ increases and this potential decreases more sharply with an increase in $l$. On the other hand, when the potential is observed with respect to $l$, one can see that the potential increases linearly with $l$ as visualized in \figurename{\ref{Vselectricb}}.

\begin{figure}[h]
	\centering
	\subfigure[]{\includegraphics[width=8cm,height=7cm]{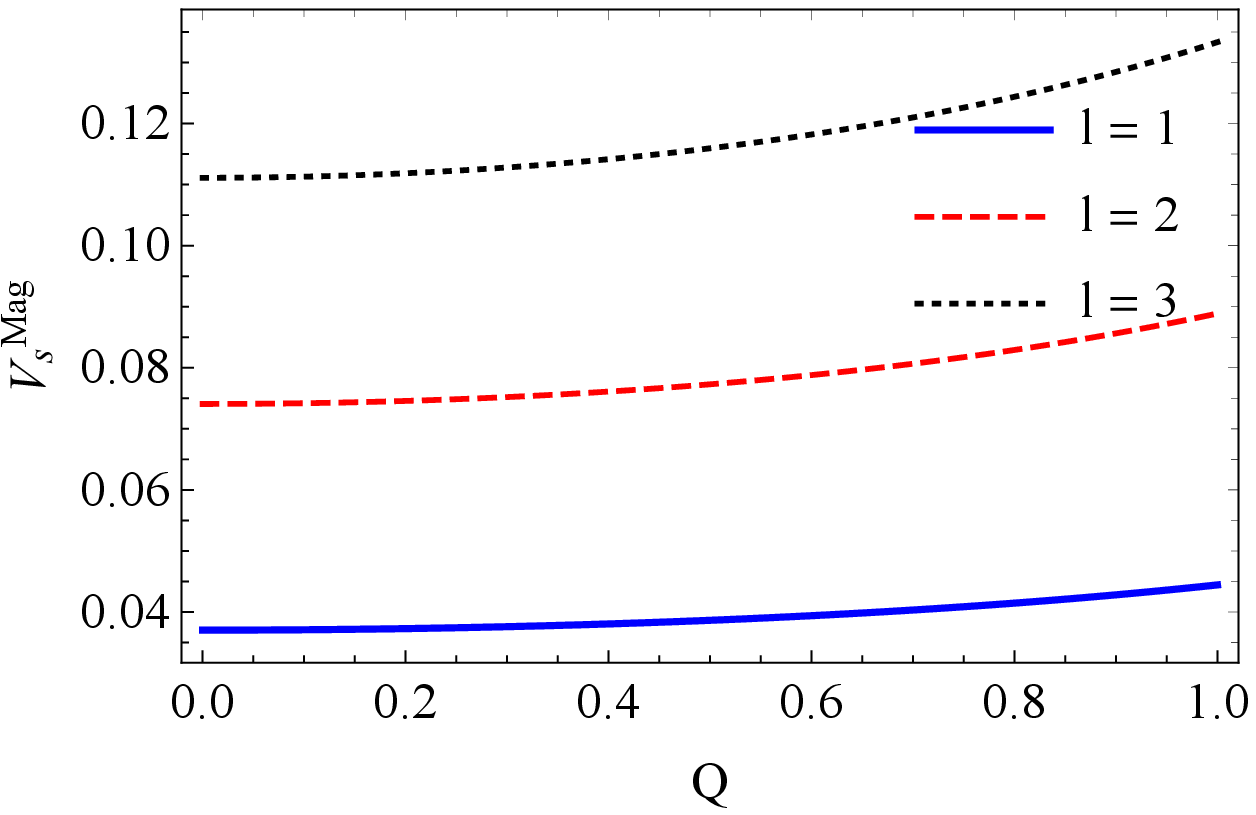}\label{Vsmagnetica}}
	\subfigure[]{\includegraphics[width=8cm,height=7cm]{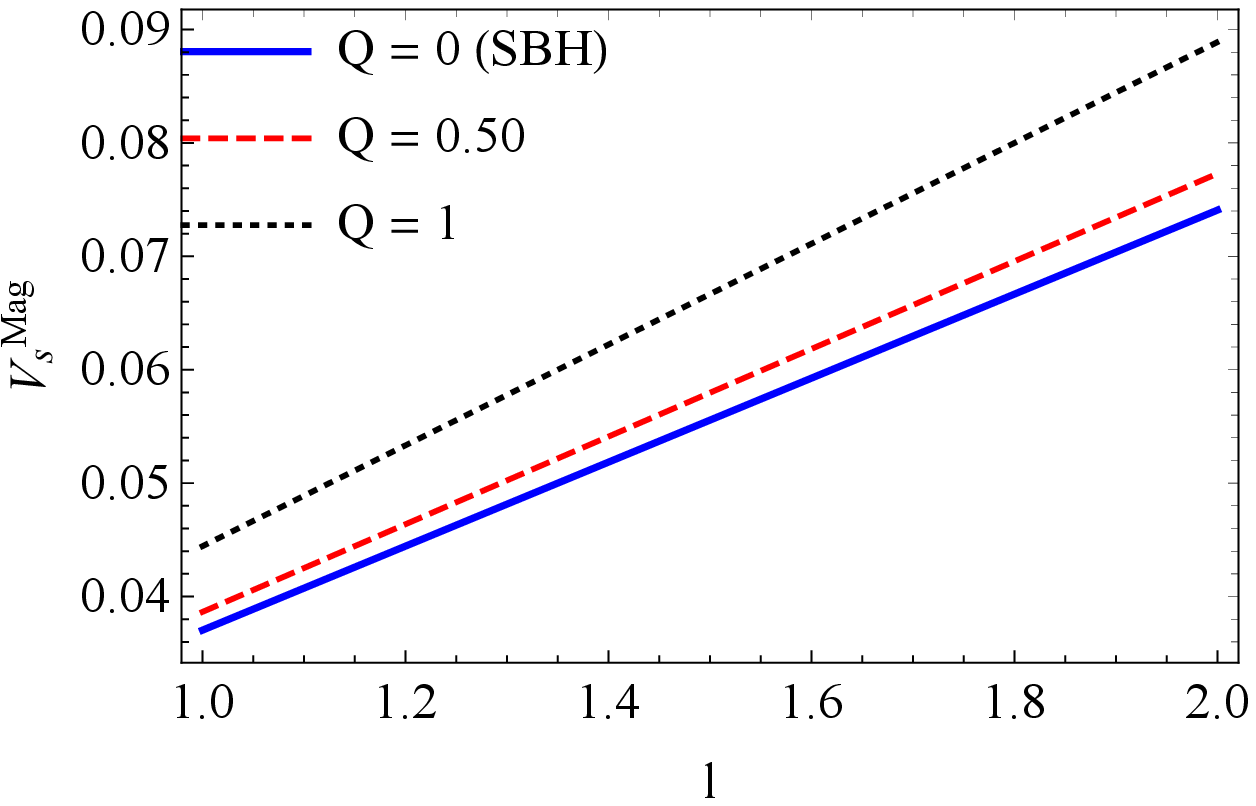}\label{Vsmagneticb}}
	\caption{The variation of scalar field potential $``V_{s}^{Mag}"$ with magnetic charge parameter$``Q"$ by varying the value of $l$ for fixed $M=1$ and $r_{c}=3M$ (see upper panel). The variation of $``V_{s}^{Mag}"$ with $``l"$ by varying $Q$ for fixed $M=1$ and $r_{c}=3M$ (see lower panel).}
\end{figure}

However, the scalar field potential for magnetic charged stringy BH is expressed as,
\begin{equation}
	V_{s}^{Mag}(r) = l (\Omega_{c}^{Mag})^{2}=l \frac{2M \left(r_{c}-2M\right)}{r_{c}^{2}\left(2M r_{c}-Q^{2}\right)}.\label{eq78}
\end{equation}

Here, the scalar field potential $V_{s}^{Mag}$ also depends upon the mass of BH ($M$), magnetic charge parameter ($Q$) and angular momentum of the perturbation ($l$). The variation of scalar field potential  with respect to magnetic charge parameter $Q$  by varying angular momentum of perturbation $l$ is presented in \figurename{\ref{Vsmagnetica}}, in which the potential increases with increasing $Q$, while \figurename{\ref{Vsmagneticb}} visualizes the variation of potential with $l$ for fixed values of other parameters. One can observed that as $l$ increases, the potential linearly increases accordingly.

\noindent The QNMs and parameters related to unstable circular null geodesics are valid in large $l$-limit $(l \to  \infty)$ and are associated with each other by a interesting relationship that established by Cardoso et al. \cite{cardoso2009geodesic} in the following form,

\begin{equation}
	\omega_{QNM}= l\Omega_{c}-i\left(n+\frac{1}{2}\right)\lambda_{Null}.\label{eq80}
\end{equation}
The angular frequency $(\Omega_{c})$ at the unstable null geodesic and the Lyapunov exponent~(reciprocal to instability timescale of orbit) for null circular geodesics $(\lambda_{Null})$ are therefore the real and imaginary parts of QNMs of the BH respectively. The Lyapunov exponent appearing in Eq.~(\ref{eq80}) may thus be interpreted as the decay rate of the unstable circular null geodesics.\\

Now, by inserting Eqs.(\ref{eq47}) and (\ref{eq49}) into Eq.(\ref{eq80}), the QNM frequencies for stringy BH with electric charge is obtained as,

\begin{equation}
	\omega_{QNM}^{Ele} =l\Omega_{c}^{Ele}-i\left(n+\frac{1}{2}\right)\lambda_{Null}^{Ele},
\end{equation}

\begin{multline}
	\omega_{QNM}^{Ele} = l \sqrt{\frac{\left(r_{c}-2M\right)}{r_{c} U^{2}}}-i\left(n+\frac{1}{2}\right) \\ \sqrt{\frac{\left(r_{c}-2M\right)^{2}}{U^{4}}\left[\frac{4M^{2}\sinh^{4}\alpha}{r_{c}^{2}}+\frac{4M \sinh^{2}\alpha U}{r_{c}^{2}}-\frac{3U^{2}}{r_{c}^{2}}+\frac{6M U^{2}}{r_{c}^{2}\left(r_{c}-2M\right)}\right]}.\label{eq81}
\end{multline}

Thus, both the real and imaginary parts of QNMs of electric charged stringy BH are dependent on \emph{electric charge parameter ($\alpha$)}.

\noindent By substituting the Eqs.(\ref{eq70}) and (\ref{eq72}) into Eq.(\ref{eq80}), we have further obtained the QNM frequencies for stringy BH with magnetic charge as follows,

\begin{equation}
	\omega_{QNM}^{Mag} =l\Omega_{c}^{Mag}-i\left(n+\frac{1}{2}\right)\lambda_{Null}^{Mag},
\end{equation}

\begin{equation}
	\omega_{QNM}^{Mag}= l \sqrt{\frac{2M \left(r_{c}-2M\right)}{r_{c}^{2}\left(2M r_{c}-Q^{2}\right)}}-i\left(n+\frac{1}{2}\right)\sqrt{\frac{3\left(r_{c}-2M\right)\left(4M-r_{c}\right)}{r_{c}^{4}}}.\label{eq82}
\end{equation}

So, in case of magnetic charged stringy BH, the real part of QNMs is dependent on magnetic charge parameter ($Q$) while imaginary part is independent of $Q$.

The above two expressions (\ref{eq81}) and (\ref{eq82}) of QNMs for stringy BHs reduce to QNMs of SBH at the eikonal limit with $\alpha=0,~ Q=0$ for $r_{c}=3M$ as below,

\begin{equation}
	\omega_{QNM}^{SBH} =l\sqrt{\frac{M}{r_{c}^{3}}}-i\left(n+\frac{1}{2}\right)\frac{\sqrt{3}M}{r_{c}^{2}}.
\end{equation}

Significantly, the expressions of QNMs for any BH spacetime in the eikonal approximation must be consisted of the real and imaginary parts which are termed as the angular frequency and parameter related to the instability timescale of the orbit~(i.e. Lyapunov exponent) corresponding to the unstable null circular geodesics respectively.

\section{Summary, Conclusions and Future Directions} 

In this work, we have performed the stability analysis of timelike as well as null circular geodesics of test particle in 3+1 dimensional spacetimes representing stringy BHs which determine the important features of these spacetimes. The main conclusions drawn from our investigations are briefly summarized as follows:
\begin{enumerate}
		\item  The proper time Lyapunov exponent $(\lambda_{p})$ and coordinate time Lyapunov exponent $(\lambda_{c})$ are derived explicitly to investigate the full descriptions of stability of timelike and null circular geodesics on the equatorial plane of both BH spacetimes. From which, we have outlined the following results:\\
	\begin{itemize}
		\item \noindent For stringy BH with electric charge, the timelike circular geodesics are stable when $\Delta<0$ such that both Lyapunov exponents are imaginary. These geodesics are unstable when $\Delta>0$ i.e. both exponents are real and they are marginally stable when $\Delta=0$ for which both exponents vanish simultaneously. Also, the null circular geodesics are unstable because the Lyapunov exponent $(\lambda_{Null}^{Ele})$ at $r_{c_{\pm}}$ is real and finite. \\ 
		
		\item \noindent For stringy BH with magnetic charge, the timelike circular geodesics are stable with $\left(r_{0}-6M\right)>0$ and $\left(2Mr_{0}-Q^{2}\right)>0$ such that $\lambda_{p}^{Mag}$ and $\lambda_{c}^{Mag}$ are imaginary. The geodesics are however unstable when $\left(r_{0}-6M\right)<0$ and $\left(2Mr_{0}-Q^{2}\right)<0$ such that both Lyapunov exponents should have a real value. \\
		
		\item \noindent The coordinate time Lyapunov exponents for magnetic charged BH are independent of magnetic charge parameter $(Q)$ for both cases i.e. timelike ($\lambda_{c}^{Mag}$) as well as null ($\lambda_{Null}^{Mag}$). 
	\end{itemize}
\item The variation of the ratio of Lyapunov exponents $(\lambda_{p}/\lambda_{c})$ for electric charged stringy BH as function of the radius of circular orbit $(r_{0}/M)$ is presented accordingly. It is observed that the ratio varies from orbit to orbit for different values of electric charge parameter $(\alpha)$ and decreases exponentially with $r_{0}/M$. On the other hand, in case of magnetic charged stingy BH, the Lyapunov exponent ratio with $r_{0}/M$ decreases exponentially for various values of magnetic charge parameter $(Q/M)$ and it has same nature as that of a SBH (i.e. $Q/M=0$) since the contribution of the magnetic charge parameter is insignificant.
	
	\item For null circular geodesics, we have computed instability exponent $\lambda_{Null}/\Omega_{c}$ to understand the instability of unstable circular orbits for both the cases of stringy BH. For electrically charged BH, it is visualized that instability of null circular orbit increases exponentially with electric charge parameter ($\alpha$). On the other hand, for magnetically charged BH, the instability exponent decreases sharply with increase in magnetic charge parameter $Q$ for $M=1$. We have further presented the variation of scalar field potential with respect to charge parameters ($\alpha$ or $Q$) and angular momentum of perturbation $(l)$ for both BH spacetimes. The scalar field potential ($V_{s}^{Ele}$) decreases with increase in $\alpha$ and this potential decreases more sharply for higher values of $l$. On the other hand, when the potential is observed with respect to $l$, it increases linearly with $l$.
	
	\item Furthermore, we have also computed the characteristic modes or QNMs those are explained by the unstable null circular geodesics. We generalize relationship between QNMs for a massless scalar field perturbation near BH in the eikonal limit and the parameters of unstable null circular geodesics. One can conclude that the real part of the complex QNMs is the orbital angular velocity (or angular frequency) $\Omega_{c}$  and the imaginary part is  related to the instability timescale of the orbit (i.e. Lyapunov exponent) calculated for unstable null circular geodesics. As a result, both parts of QNMs ($\omega_{QNM}^{Ele}$) are dependent on the electric charge parameter ($\alpha$). On the other hand, the real part of QNMs ($\omega_{QNM}^{Mag}$) depends on magnetic charge parameter ($Q$)  while the imaginary part is independent of $Q$.
	
	\item All the results obtained here are easily reduced to SBH case in the prescribed limit ($\alpha=0$ and $Q=0$) \cite{pradhan2016stability}.\\
	It would further be interesting to study the quantum gravity effects on unstable circular orbits for these spacetimes using the Lyapunov exponent as studied by Dasgupta~\cite{dasgupta2010quantum} for SBH. In near future, we intend to investigate the stability analysis of geodesics around some rotating BH spacetimes those emerged in GR and other alternative theories of gravity.
	
\end{enumerate}


%
%


\section*{\normalsize Acknowledgments}
{\normalsize We would like to express our gratitude to Radouane Gannouji and Parthapratim Pradhan for useful discussions. One of the authors SG thankfully acknowledges the financial support provided by University Grants Commission (UGC), New Delhi, India as Junior Research Fellow through UGC-Ref.No. {\bf 1479/CSIR-UGC NET-JUNE-2017}. HN would like to thank Science and Engineering Research Board (SERB), India for financial support through grant no. {\bf EMR/2017/000339}. The authors also acknowledge the facilities at- ICARD, Gurukula Kangri (Deemed to be University) Haridwar.}
\appendix
\renewcommand{\theequation}{A-\arabic{equation}}
\setcounter{equation}{0}  
\section*{Appendix-A: Relation between Lyapunov Exponent and Radial Effective Potential}  
In the viewpoint of stability analysis of any dynamical system, the concept of Lyapunov exponent has been used widely. Let us consider an observed trajectory denoted by $x(t)$, which is the solution of an equation of motion in $d$-dimensional phase space given by~\cite{sano1985measurement},
\begin{equation}
	\frac{dx}{dt}=F(x).\label{eq1}
\end{equation}
Now, if one simply apply a small perturbation $\xi(t)$ on $x(t)$ in order to calculate the stability of trajectory given as,
\begin{equation}
	x(t)= x_{0}+\xi(t),\label{eq2}
\end{equation}
where, $x_{0}$ is a fixed point at t=0.
By inserting Eq.(\ref{eq2}) into Eq.(\ref{eq1}), we have,
\begin{equation}
	\frac{d\xi}{dt}=F(x_{0}+\xi).\label{eq3}
\end{equation}
The Taylor series expansion of Eq.(\ref{eq3}) about $x_{0}$ and linearizing it about 
certain orbit leads to,
\begin{equation}
	\frac{d\xi}{dt}= T(x(t))\xi ,\label{eq4}	
\end{equation}
where,
\begin{equation}
	T(x(t))=\frac{\partial F}{\partial x},\label{eq5}
\end{equation}  
is known as the linear stability matrix. The eigenvalues of the Jacobian matrix $T(x(t))$ 
are known as characteristic exponents or Lyapunov exponents associated with $F$ at the fixed point 
$x=x_{0}$~\cite{pradhan2016stability}.\\

The solution of the linearized Eq.(\ref{eq4}) can be expressed in the following form,
\begin{equation}
	\xi(t)= \Phi(t) \xi(0),\label{eq6}
\end{equation}
where, $\Phi(t)$ is the evolution matrix or operator that maps tangent vector $\xi(0)$ to $\xi(t)$.

The mean exponential rate of expansion or contraction in the direction of $\xi(0)$ on the trajectory passing through trajectory $x_{0}$ is given by the eigenvalues of $\Phi(t)$  as defined below,
\begin{equation}
	\lambda= \lim_{t\to\infty}\frac{1}{t} ln \frac{\parallel\xi(t)\parallel}{\parallel \xi(0)\parallel},\label{eq7}
\end{equation}
here,  $\parallel..\parallel $ implies a vector norm and the quantity $``\lambda"$ is called the principal Lyapunov exponent.\\

\noindent Let us derive the generalized relation between second derivative of the square of the radial component of the four-velocity and the Lyapunov exponent. For any static, spherically symmetric spacetime with metric component~$(g_{ii})$, the necessary Lagrangian of a test particle motion in the equatorial plane~($\theta=\frac{\pi}{2}$) is described as,
\begin{equation}
	\mathcal{L}= \frac{1}{2}\left[ g_{tt} \dot{t}^{2}+ g_{rr} \dot{r}^{2} + g_{\phi\phi}\dot{\phi}^{2}\right]
	.\label{eq8}
\end{equation}
The generalized momenta associated with Lagrangian is given by,
\begin{equation}
	p_{q}= \frac{\partial	\mathcal{L}}{\partial \dot{q}},\label{eq9}
\end{equation}
where, $q \equiv (t,r,\phi)$ and over dot $(.)$ denotes the differentiation with respect to the 
proper time~($\tau$).

The well-known Euler-Lagrange equations of motion is given by,
\begin{equation}
	\frac{d}{d\tau}\left(\frac{\partial	\mathcal{L}}{\partial\dot{q}}\right)
	=\frac{\partial	\mathcal{L}}{\partial q}.\label{eq10}
\end{equation}
On substituting Eq.(\ref{eq9}) into Eq.(\ref{eq10}), one can have an another form of equation 
of motion as follows,
\begin{equation}
	\frac{dp_{q}}{d\tau}=\frac{\partial	\mathcal{L}}{\partial q} .\label{eq11}
\end{equation}
Thus, from Eqs.(\ref{eq9}) and (\ref{eq11}), considering a two dimensional phase space of the form 
$ X_{i} = (p_{r},r)$, we obtain,
\begin{equation}
	\dot{p_{r}}= \frac{\partial	\mathcal{L}}{\partial r}~~~~ \& ~~~~\dot{r}= \frac{p_{r}}{g_{rr}} .\label{eq12}
\end{equation}
Now, by linearizing the equation of motion~(\ref{eq12}) about an orbit of constant radius $r=r_{0}$, the evolution matrix can be expressed as,
\begin{equation}
	K= \begin{pmatrix}
		0 & \frac{d}{dr}\frac{\partial	\mathcal{L}}{\partial r}\\
		\frac{1}{g_{rr}} & 0
	\end{pmatrix}.\label{eq13}
\end{equation}
Therefore, the eigen values of the evolution matrix along the circular orbits are evaluated as,
\begin{equation}
	\lambda^{2}= \frac{1}{g_{rr}}\frac{d}{dr}\frac{\partial	\mathcal{L}}{\partial r},\label{eq14}
\end{equation}
which are so-called the principal Lyapunov exponents.

The equations of motion~(\ref{eq10}) for radial coordinate will lead to an equation of the form,
\begin{equation}
	\frac{d}{d\tau}\left(\frac{\partial	\mathcal{L} }{\partial\dot{r}}\right)= 
	\frac{\partial	\mathcal{L}}{\partial r},\label{eq15}
\end{equation}
which yields to,
\begin{equation}
	\frac{\partial	\mathcal{L}}{\partial r}=\frac{1}{2g_{rr}}\frac{d}{dr}{g_{rr}^{2}\dot{r}^{2}}.\label{eq16}
\end{equation}
Finally, the principal Lyapunov exponent~(\ref{eq14}) can be rewritten as,
\begin{equation}
	\lambda^{2}= \frac{1}{g_{rr}}\frac{d}{dr}\left(\frac{1}{2g_{rr}}
	\frac{d}{dr}{g_{rr}^{2}\dot{r}^{2}}\right).\label{eq17}
\end{equation}
The condition for circular geodesics at $r=r_{0}$ is defined as,
\begin{equation} 
	\dot{r}^{2}= (\dot{r}^{2})^{'}=0,\label{eq18}
\end{equation}
here and throughout the paper, the prime ($'$) stands for the differentiation with respect to  $r$.\\

Using Eq.(\ref{eq17}) with considering the condition of circular orbits (\ref{eq18}), one can obtain the proper time Lyapunov exponent as expressed below \cite{pradhan2016stability},
\begin{equation}
	\lambda_{p}=\pm\sqrt{\frac{\left(\dot{r}^{2}\right)^{''}}{2}}= \pm\sqrt{\frac{V(r)^{''}}{2}},\label{eq19}
\end{equation}
and the coordinate time Lyapunov exponent can be expressed as~\cite{cardoso2009geodesic,pradhan2016stability},
\begin{equation}
	\lambda_{c}=\pm\sqrt{\frac{\left(\dot{r}^{2}\right)^{''}}{2 \dot{t}^{2}}}= \pm\sqrt{\frac{V(r)^{''}}{2 \dot{t}^{2}}}.\label{eq20}
\end{equation}
We will drop the sign ($\pm$) from Eqs.(\ref{eq19}) and (\ref{eq20}) in our calculations throughout the work, i.e. only positive Lyapunov exponent is to be considered. 
In case the Lyapunov exponent $\lambda_{p}$ (or $\lambda_{c}$) is real then the circular orbit is unstable however the circular orbit is stable for imaginary nature of the $\lambda_{p}$ (or $\lambda_{c}$). It is marginally stable when $\lambda_{p}$ (or $\lambda_{c}$) is zero~\cite{pradhan2016stability}.
\bibliographystyle{unsrt} 
\bibliography{shobitrefLYEX} 

\end{document}